# AC Conductivity in Boron Doped Amorphous Conducting Carbon Films on the Insulating Side of MI Transition

*P.N.Vishwakarma\**
*Department of Physics*
*Indian Institute of Science, Bangalore*
*India – 560012*

## Abstract:

Boron doped amorphous conducting carbon films show MI transition induced by doping.  In this paper we discuss the ac conductivity of the films, which lie on the insulating side of MI transition. The ac conductivity data is analyzed as a function of frequency as well as temperature for both real as well as imaginary part of the conductivity data. The ac conductivity of these samples shows enhanced interaction effect at low temperature. The conduction mechanism at high frequency and at low temperature is due to the tunneling mechanism. At intermediate temperatures and at moderate frequencies, the conductivity data is in good agreement with extended pair approximation modified for interaction effect. At high temperature and at low frequencies, the mechanism of ac conductivity is similar to that of dc conduction. The conductivity data for the insulating samples near the boundary of MI transition were analyzed using an RC element model. While fitting the impedance data, a new type of constant phase element (inductive instead of capacitive) was found to be appropriate. At present the origin of this inductive CPE is not is clear but we believe its origin is in the electron-electron interaction, which makes any attempt of applied frequency a slower response. Our ac conductivity results are well explained by tunneling models rather than hopping models.

\* email : prakash@physics.iisc.ernet.in

# 1. INTRODUCTION

AC conductivity in disordered materials has been a field of interest for the scientists due to their unexpected behaviour as compared to their crystalline counter parts. While ordered solids show no frequency dependence of their conductivity at frequencies below phonon frequencies, disordered solids are characterized by ac conductivity that varies as an approximate power law of frequency. Generally the conductivity shows a power law in frequency as well as temperature, $\sigma \propto \omega^s T^n$. The exponent '$s$' is usually less than but close to one. As the frequency goes to zero the conductivity becomes frequency independent. This universality is observed in many disordered materials like amorphous semiconductors [1,2,3], polymers [4], conducting disordered solids like glasses [2,5]. A number of models were put forward to explain this universal behaviour of ac conductivity phenomenon in disordered solids. All these models can be broadly categorized in two categories, namely microscopic and macroscopic models. The microscopic model [6] assumes disorder on atomic length scale and the conductivity is due to either hopping or tunneling through the localized states. Whereas the macroscopic model [7] assumes disorder on length scales large enough that a local conductivity may be defined and the conductivity is explained on the basis of either effective medium approximation (EMA) or percolation path approximation (PPA). In this paper we will be discussing ac conduction of boron doped amorphous conducting carbon films in the insulating side of MI transition in the framework of available theoretical models. These films show a MI transition, which depends upon the preparation temperature and on the boron concentration in the carbon matrix. Also for the poorly conducting films, coulomb gap due to interaction effect is also observed. The class of amorphous carbon films, which has been discussed here, are rich in sp2 hybridized carbon atoms. These sp2 carbon atoms are responsible for the high conducting nature of these films. The conductivity of these films very much depends on the preparation temperature. The films prepared at $900^0$C are relatively more conducting (metallic) than that prepared at $700^0$C (insulating). The insulating nature of the films prepared at $700^0$C are not due to high sp3 concentration, rather it is induced by disorder in the system. To estimate the sp2/sp3 ratio, Raman analysis was done on these samples. However due to relatively large scattering cross-section for sp2 than that of sp3, the results were not very conclusive [8]. So the exact percentage of sp3 carbon atoms in the sp2 background is not known. However the XRD data on these films show increasing graphitization (or ordering) with increasing preparation temperature. So, it is believed that the transition in the film from metallic to insulating state, induced by changing the preparation temperature is due to highly disordered nature of the film. The observed MI transition can be very well understood in terms of density of states. The conduction mechanism in the system is mainly governed by the states near the Fermi level. If the disorder in the system is very low so that the band edge $E_C$ lies far below the Fermi level, than the system will behave metallic because the states governing the conduction mechanism (near the Fermi level) are extended in nature. However the same system can be driven to an insulating state by increasing the disorder in the system so that the band edge crosses the Fermi level and all the states near the Fermi level are localized.

When these amorphous carbon films are doped with boron, the boron atoms take the atomic site previously taken by the carbon atoms. The pi electron of carbon, which was responsible for electrical conduction in undoped carbon films, is missing at the site replaced by boron. Eventually the Fermi level in the case of boron-doped carbon films will come down due to low density of conduction electrons. If the carbon film is sufficiently doped to bring down the Fermi level below the band edge, the system will go to an insulating state. However it is not possible to dope any system without introducing any disorder. So in actual case the Fermi level will come down due to decrease in the density of conduction electrons and the band edge will move up due to lattice disorder introduced by boron in carbon network due to mismatch in size. These two processes are simultaneous and it is difficult to separate them. From the conductivity data, though it may look like an increase in sp3 nature due to boron doping, actually it is not so. The decrease in conductivity of boron doped carbon films are entirely due to less conduction electron density. The density of electron states calculated form the experimental data is indeed showing a decreasing trend with increasing boron concentration in the carbon network. Also if the sp3 nature of carbon is increasing, the carbon films should become transparent to the visible light. However this was not observed. So from this we can confidently say that the MI transition in amorphous carbon films brought by the boron doping is not due to increasing sp3 content but it is due to decreasing conducting pi electrons. Since, boron carbide is one of the hardest crystalline phase of Boron and Carbon, the possibility of nano-crystallites of boron carbide embedded in the background amorphous phase was examined using microhardness measurement and transmission electron microscopy (TEM). Microhardness measurements did not give any indication of hard phase present in the film. On contrary, the hardness of the films detoriated when boron was doped in the carbon network. The decrease in the hardness of the film is another indication of less density of sp3 bonding in the film. The microstructure of the boron doped carbon film was examined using TEM. The TEM data does not show any boron particles embedded in the film. The film was homogeneous throughout the examined region. So we can say that the disorder in the film is on the length scale much smaller than that proposed for macroscopic models. The X-ray and selected area electron diffraction (SAED) of the films reveals a complete amorphous nature of the films.

Though there are few reports of ac conductivity in insulating side of samples showing MI transition **[9]**, extensive work in this field is still very less. Generally only the real part of conductivity data is analyzed while imaginary part is ignored. However we feel this is not correct. For applicability of any model, it should be tested for both real and imaginary part of conductivity. So here we will be discussing the ac conduction in boron doped amorphous conducting carbon films in the framework of available theories for real as well as imaginary part of conductivity as a function of frequency as well as temperature.

## 2. EXPERIMENTAL DETAILS

The boron doped amorphous carbon films were prepared by pyrolysis assisted CVD technique using two-zone furnace method. The details of preparation and structural characterization are given elsewhere **[10]**. The atomic percentage of boron in these carbon films was 25%(CB700M), 18%(CB700M2) and 10%(CB700M4). For the notation of sample CB700M2, C and B stands for carbon and boron, 700 is the pyrolysis temperature in $^0$C and M2 stands for half molar, M4 quarter molar and M one molar boric acid solution initially taken as precursor. The carbon films having more boron content were less conducting. AC conductivity for all the samples were measured from 1Hz to 102KHz in the temperature range of 300 K to 1.2 K using liquid helium cryostat supplied by Janis Corp, USA. The electrical contacts to the sample were made using conducting silver epoxy in accordance to standard four-probe technique. Impedance measurements were performed using SR 830 lock-in amplifier (Stanford Instruments). Lock-in amplifier does not directly give the impedance value but it can be calculated by using the following formula.

$$Z = \frac{V_S}{I_S} = \frac{V_S}{V_R} \times R \tag{1}$$

$$= R \times \left( \frac{V_S{}' + jV_S{}''}{|V_R|} \right)$$

$$= R \times \left( \frac{V_S{}' + jV_S{}''}{\sqrt{(V_R{}')^2 + (V_R{}'')^2}} \right) \tag{2}$$

So,

$$Z' = R \times \frac{V_S{}'}{\sqrt{(V_R{}')^2 + (V_R{}'')^2}} \tag{3}$$

$$Z'' = R \times \frac{V_S{}''}{\sqrt{(V_R{}')^2 + (V_R{}'')^2}} \tag{4}$$

where $V_S$ and $V_R$ are the voltages across the sample and standard resistor R=100 ohm connected in series with the sample and the lock-in amplifier. The purpose of the standard resistor is to know the current passing through the sample, as lock-in amplifier does not give constant current. Voltage drop across the standard resistor is purely real, but due to capacitance effect there will also be imaginary current in the circuit, which will give rise to imaginary voltage across the resistor. So to know the total current flowing in the circuit, one has to take the modulus of the voltage drop across the standard resistor. Prior to experiments on the carbon samples, the experimental setup was calibrated using standard resistors varying from 10 ohm to 20 mega ohm. This was done for the entire temperature as well as frequency range i.e., 4.2K - 300K and 1Hz - 102kHz. So before collecting the data from the samples proper care was taken so that no out of phase component is arising from the circuit itself. So the data presented here is only from the sample and the experimental noise is negligible.

The ac conductivity spectra for all the samples in general can be shown to have behaviour as shown in figure 1. The real and imaginary conductivity spectra for samples C700, CB700M4, CB700M2 and CB700M at four different temperatures 4.2 K, 31 K, 77 K and 300 K are shown in figures 2 to figure 5. The temperature dependence of ac conductivity for the same samples at four different frequencies 1 kHz, 10 kHz, 50 kHz and 100 kHz are depicted in figure 6 to figure 9.

## 3. RESULTS AND DISCUSSION

The classification of metallic and insulating samples was done using reduced activation energy 'W' ($W(T) = -\partial(\log \rho) / \partial(\log T)$) plots as suggested by Zabrodskii [11]. According to this classification scheme if the plot of W versus T in log-log scale is having negative slope, it is termed insulating where as if the slope is positive, the system is considered metallic. For our samples we observed a transition from positive to negative slope either by decreasing the preparation temperature or increasing the boron content of the carbon film. The MI transition in these films is also supported by the fact that in metallic films, the conduction mechanism was in agreement with weak localization theory whereas the insulating films were following hopping models described by Mott and Efros and Shklovskii.

Recently the macroscopic technique adopted by Dyre and Schroder [12] has got lot of attention in understanding the ac conductivity behaviour in disordered solids. However group IV elements have been exception to this universality and the ac conductivity behaviour of this group of materials is still under debate. In such solids the electrical conduction takes place via classical hopping over the potential barrier or quantum mechanical tunneling of charge carriers through the barrier. The hopping process is due to thermal activation of charge carriers, and hence they are relatively high-temperature phenomena than quantum mechanical tunneling.

We will be discussing the ac conductivity data for boron doped amorphous carbon films having varying boron concentration and prepared at different temperatures. The analysis is done on the basis of available models. Depending on the conductance spectra, we have done the analysis in two different ways. The highly resistive samples were analyzed using the universality criterion. Universality was not observed even at highest frequencies for the samples, which were relatively more conducting. For such samples, the data was analyzed using impedance spectra.

A. Real Part

First let us start with the real part of ac conductivity. Here we have divided the entire spectra into four regimes depending upon the frequency response of the conductivity.

## *1. Regime I*

At low frequencies the conductivity value is roughly the same as the dc conductivity (red solid curve in figure 1). In this time scale (inversely proportional to the frequency applied), the applied electric field is not able to perturb the hopping conduction mechanism of carrier electrons. So in this frequency range the electrons see a uniform electric field during the time of flight. Hence the conductance is the dc value. Like conductance spectra, the temperature dependence of conductivity graph also can be divided into four regimes corresponding to conductance spectra (figure 6). At low frequencies and high temperatures, the ac conductivity curve merges into the dc conductivity. At high temperatures, experimental data taken at different frequencies tend to collapse in a single curve, corresponding to the dc conductivity (regime I). In this regime the conduction mechanism is same as that for dc conduction i.e., via hopping of electrons from one localized site to another.

## *2. Regime II*

As the frequency is increased the conductivity does not change until a critical frequency ($\omega_0$). After this critical frequency the conductivity increases slowly but nonlinearly (regime II). For few samples the conductance data is limited only to regime II and I and show no universality. Such samples can be understood by analyzing them in terms of appropriate R-C circuit. The simplest R-C element is R and C parallel to each other. So the analysis for such samples was done for impedance spectra rather than conductivity. In regime II, with an increase in frequency, an increasing number of capacitor admittances become numerically larger than the admittance of their resistor partner. Whenever this happens for a link (an RC element) it is termed "affected". The average resistor current changes only insignificantly as long as none of the affected links are on the fat percolation cluster (which carries almost all currents) **[12]**. When frequency is increased further, at some point, links on the fat percolation cluster do become affected. The first of these are the bottlenecks, the links with the largest resistors. From there on, as the frequency is further increased, more and more links on the fat percolation cluster are affected – the node potentials on the cluster change and so do the resistor currents: The conductivity becomes frequency dependent. The frequency-marking onset of ac conduction is roughly proportional to the dc conductivity, because both are roughly proportional to the bottleneck admittance. Barton **[13]**, Nakijama **[14]** and Namikawa **[15]** independently verified for many materials that $\omega_0 = p\sigma_{dc} / (\varepsilon_0 - \varepsilon_\infty)$ where p is a constant close to one, $\varepsilon_0$ and $\varepsilon_\infty$ are the static and high frequency dielectric constants. This equation is also known as BNN relation. For the temperature dependence in this regime, an additional contribution to the power law dependence of ac conductivity is observed. In this frequency regime, electrons have already jumped over the potential barriers of their wells but a stationary behaviour, i.e. a dc conductivity value, for long range electron transport has not been achieved.

### 3. Regime III

In regime III the conductivity in log scale increases linearly with slope nearly equal to two. In microscopic model this quadratic behaviour is explained as breakdown of pair-approximation. Generally it is assumed that at high frequency, the relaxation is confined entirely to pairs of sites, and multiple site participation is excluded. However, the pair approximation has been found to be good only at sufficiently high frequencies. The pair approximation breaks down at low frequencies, where a percolation path must be established throughout the sample in the dc limit. These percolation paths should not be confused with that of percolation model. These percolation paths are just a continuous path for electrical conduction that is responsible for dc conductivity. Butcher and coworkers developed a so-called "extended pair approximation" (EPA) [16] in which the effect on the relaxation rate for a given pair of sites of all the other sites in the network is taken into account in an averaged way. This effect was neglected in pair approximation where the hopping or tunneling was restricted to a pair of potential wells. Using EPA approach, the correction to the real part of ac conductivity for low frequency was found to be proportional to $\omega^2$ [17].

The regime III corresponds to an increase in conductivity with a decrease in temperature, before attaining a constant value. Using extended pair approximation and in the absence of interaction, it was suggested that at low frequency (near to threshold frequency), conductivity goes as $\omega^2$ and is proportional to '1/T', i.e. the slope of the curve shown in the figure should be $-1$ [17]. However we found the slope to be -2.6. Recently it was proposed by Murugavel et.al that whenever interaction effect is dominant, interaction correction to the temperature dependence of conductivity should be $\propto T^{-2.3}$ [18]. Our observed value of the slope $-2.6$ is quite close to that observed by Murugavel et.al of $-2.3$. Baranvoskii and Cordes [19] carried out analytical calculations on the dynamics of a single particle using a percolation approach and they analyzed in detail the scaling properties of the conductivity spectra. Thereby they found the temperature exponent to be -2.3. This analytical result was qualitatively confirmed by recent computer simulations carried out by Schroder and Dyre [20,12]. Recently, Monte-Carlo simulations on the particle dynamics in the RBM model, taking into account long-range coulomb interactions between the mobile electrons shows that the exponent increases with increasing strength of coulomb interactions [21].

### 4. Regime IV

As the frequency is increased, i.e. when the pair approximation is approached, $\sigma(\omega)$ changes from $\omega^2$ dependence to $\omega$ dependence (regime IV). Such roll-off behaviour at low frequencies has also been observed in amorphous Si and Ge [3]. Since $\sigma' \propto \omega$ results in a dielectric loss nearly independent of frequency ($\varepsilon'' = \sigma'/\omega$), such behaviour is usually known as nearly constant loss (NCL). Though the existence of this NCL was suggested more than 25 years ago and subsequently verified in other materials [20], the origin of this NCL in various materials is still under debate. In our samples, it seems to be linear dependence of conductivity on frequency. Careful analysis shows the exponent 's' to be slightly more than one ($\approx 1.05$). We found this super-linear value of 's' to be consistent in all the samples showing linear dependence at low temperature. Such value of 's $\approx 1.05$' has also been observed in many other materials [23]. The only theoretical model, which deals with superlinear frequency exponent is, relaxation via tunneling when interactions among the electrons are dominant [3, 6, 24]

$\sigma'(\omega) \propto \omega^{s'}$ ; where $s' = 1 + \dfrac{1}{\ln(1/\omega\tau_{0t})}$

$\sigma'' \propto \omega^{s''}$ ; where $s'' = 1$

and the corresponding temperature exponent,

$n' = n'' = 0$

In terms of microscopic theory, the exciting field changes the relative environment of a pair of centers and causes transition between them governed by the intrinsic relaxation time $\tau$ of the pair. The resultant loss is dominated by pairs having relaxation times $\tau$ around $\omega^{-1}$. So in the limit of high frequency, transitions are taking place mostly from those sites for which relaxation times are very small i.e. only from neighbouring sites.

Temperature dependence of ac conductivity (real part): The NCL regime is characterized by almost negligible temperature dependence. This regime is found to dominate the conductivity curves up to the lowest temperatures. In most of the materials, regime III is not seen and there is direct transition from regime II to regime IV (i.e. from non-integer power law to linear dependence). This linear frequency dependence provides a criterion to determine, at each frequency, a crossover temperature where the constant loss term becomes the dominant contribution to the ac conductivity. Dominant electron-electron interaction was also observed in dc conductivity measurements on these samples where the conductivity changes from Mott VRH to ES VRH at low temperature **[10]**. Also for the samples showing NCL behaviour, coulomb gap ( ~ 2meV) was observed.

B. Imaginary Part:

Regarding the imaginary part of the conductivity, the general form of conductivity spectra obtained for our samples are shown in the figure 1 (blue dashed curve). In the figure only the magnitude of imaginary part of the conductivity is plotted. In actual case the imaginary part of conductivity will be lagging behind the real part by a phase difference determined by the circuital analogue of the sample and hence it will be in negative scale. Unlike real part, the imaginary part of conductivity dependence on frequency as well as temperature is not very much different in regime II and I. So here we will be discussing only three regimes i.e., regime I and II, regime III and regime IV.

*1. Regime I and II*

At low frequency the log-log plot shows a linear decrease in conductivity with increasing frequency (figure 1 with dashed blue curve). This linearity goes up to a critical frequency $\omega_0$. The above-mentioned theoretical model predicts the frequency dependence of conductivity only beyond $\omega_0$ i.e., the frequency at which frequency dependent hopping length becomes comparable to the dc hopping length. So below $\omega_0$ the contribution to the imaginary conductivity is due to increasing admittance of the capacitors with increasing frequency. However this capacitor admittance is still lower than the admittance of their counter resistors, hence not affecting the potentials at the junction nodes that are responsible for the real conductivity. So the real conductivity shows frequency dependence only after $\omega_0$ and is almost unaffected till $\omega_0$, whereas the imaginary part of the conductivity shows a strong dependence with frequency. Analysis of such data using the electrical circuit analogue of R-C element, gives a more satisfactory explanation. The imaginary part of conductivity for a R-C element, whose real part is similar to that shown in figure 1, is proportional to $\omega + (1/\omega)$. So the low frequency region is dominated by $1/\omega$ term whereas high frequency data is dominated by $\omega$ term. Such behaviour has also been observed earlier **[25, 26]**. Low frequency data for our samples is indeed showing $1/\omega$ behaviour.

As discussed earlier, the conductivity for slightly conducting samples were analyzed by comparing the impedance data to that of an equivalent RC element. The fitting was very good. The values of the macroscopic resistance and capacitance for respective samples are tabulated in table I. These values of resistance and capacitance are not microscopic but are overall values. From these values, one can make out that the capacitance value is not changing much for a given set of samples and its value of the order of $10^{-2} \mu F$ where as the resistance value is changing remarkably from sample to sample. This is consistent with the models that all the capacitors are same whereas the resistors have a wide distribution **[12]**.

## *2. Regime III*

After reaching minima at $\omega_0$, the imaginary conductivity changes its slope and again starts rising with slope $\sim 1.8$ i.e. $\sigma'' \propto \omega^{1.8}$. This slope is retained for regime III and IV, where the real part of conductivity shows a change in slope from regime III to IV. The theoretical prediction for frequency dependence of imaginary part of the conductivity is linear. Sometimes such non-integer power of frequency is explained using constant phase element (CPE). It is an empirical impedance function of the type

$Z_{CPE} = A(i\omega)^{-\alpha}$

where, A and $\alpha$ are frequency-independent parameters which usually depend on temperature and $0 \le \alpha \le 1$. If $\alpha = 1$, it describes an ideal capacitor and for $\alpha = 0$ an ideal resistor. It is generally thought to arise from the presence of inhomogeneties in the electrode-material system, and it can be described in terms of a (non-randomalizable) distribution of relaxation times or it may arise from non-uniform diffusion whose electrical analog is an inhomogeneously distributed RC transmission line.

For the samples, which are showing NCL, behaviour only at lowest temperature and high frequency and not at any other temperature and frequency, the impedance spectra is fitted to a R-C element circuit. An extra term $A\omega^\alpha$, in addition to the terms due to a resistor-capacitor link was required for fitting the impedance spectra for these samples. This extra term is similar to CPE (Constant Phase Element) except the exponent $\alpha$, which is positive in this case. This extra correction to the data fitting is analogous to inductive and the exponent is less than one ($\alpha \approx 0.8$). As per our information, such non-integer inductive correction has not been observed. At present it is difficult to comment on the origin of this inductive term, but we believe its origin is in the dominant interaction among the electrons. Literally inductive response means, it will oppose the applied ac signal and allow only dc to pass through it. Here also, when NCL is reached, the interaction among the electrons is dominant. So any attempt of an electron to hop or tunnel will be a collective effect rather than an individual effect. The dominant interaction among the electrons makes this a restrictive and slow response.

In regime III, the imaginary part of conductivity increases almost linearly with slope $\approx 4.5$ (in log-log scale) up to a critical temperature. As per our information at present there is no theoretical model, which predicts such temperature dependence for imaginary part of conductivity. Hence it is difficult for us to comment on the mechanism responsible for electrical conduction giving rise to temperature exponent of 4.5 before attaining a constant NCL.

### 3. Regime IV

Since the frequency dependence of conductivity is same in regime III and IV, the same discussion will hold for regime IV regarding the frequency dependence of conductivity except for the highly resistive samples. The resistive samples show again a change in slope after reaching a maximum value at high frequency. Such change in slope has also been observed in disordered ionic solids [26]. But at present, the mechanism giving rise to this behaviour is not clear.

Temperature dependence of ac conductivity (imaginary part): According to the theory, the real as well as imaginary conductivity should show temperature independent conductivity. As can be seen in the figure, the imaginary part of conductivity is following a similar trend as the real part except the temperatures at which the conductivity is independent of temperature. Though the imaginary part of conductivity is showing conductivity independent of temperature at low temperature for CB700M (figure 9) and CB700M4 (figure7), it is not seen for other samples. The other samples are yet to reach the temperature where the conductivity goes independent of temperature. Probably at still lower temperature, they will also show conductivity independent of temperature. Low value of temperature exponent was also observed by Hauser while working in $As_2Te_3$ **[27].**

## 4. CONCLUSIONS:

The ac conductivity in boron doped amorphous conducting carbon film is well explained using electrical circuit analogue of the film for the less resistive samples. The highly resistive samples show a universal power law and the data is explained on the basis of microscopic model. At low temperature and high frequency, the real part of conductivity is directly proportional to the applied field whereas the imaginary part is showing nearly quadratic behaviour in frequency. The frequency exponent of the real part of conductivity in this frequency and temperature range is slightly more than one, which is an indication of interaction effect on tunneling assisted conduction process. Also in this frequency and temperature range, both real and imaginary conductivity is independent of temperature, confirming the tunneling assisted conduction in these samples. The quadratic dependence of conductivity on frequency near the threshold frequency $\omega_0$ and the temperature dependence near the threshold are in good agreement with EPA modified with interaction effect. Below this frequency and temperature range, the frequency exponent is non-integer. This can be explained using hopping model, but we find analyzing the impedance data using an appropriate RC element is more satisfactory. While analyzing such data using impedance analysis, use of an inductive CPE along with R and C gives a better fitting of the experimental data. The origin of this CPE is not very clear, but we believe its origin is in the interaction among the electrons. The temperature dependence of conductivity is in good agreement with that predicted and observed in the presence of electron-electron interaction. In a nutshell, we can say that analyzing the impedance data using appropriate RC element, for samples close to MI boundary, give a satisfactory fitting. For the samples lying deep inside the insulator side of MI boundary, the tunneling theory modified by taking into account interaction effects is in good agreement with the data. In the intermediate temperature and frequency near to the threshold, the EPA theory modified by Murugavel et.al is in close agreement with the experimental data.

## AKNOWLEDGEMENTS:

The author would like to thank Prof. S.V.Subramanyam for helpful discussions while preparing the manuscript. I would also like to thank S.Sarangi for extensive discussions regarding the experiment.

Table I: The respective values of R (in ohms) and C (in $\mu F$) for various samples, found by fitting the impedance spectra to relation corresponding to an RC element. Also is given the coefficient and exponent of the inductive CPE element discussed in the text. For samples showing universality, the value of frequency exponents are also given.

| Samples | 300K | 77K | 31K | 4.2K |
|---------|------|-----|-----|------|
| C700 | R = 42<br>C = 0.047 | R = 127<br>C = 0.065 | R = 653,<br>C = 0.034<br>A = $2.8 \times 10^{-3}$<br>$\alpha = 0.8$ | $s' \approx 1.02$;<br>$s'' \approx 1.83$ |
| CB700M4 | R = 191,<br>C = 0.0443 | R = 1108,<br>C = 0.050<br>A = $8.8 \times 10^{-3}$<br>$\alpha = 0.79$ | $s' \approx 1.05$;<br>$s'' \approx 1.84$ | $s' \approx 1.01$;<br>$s'' \approx 1.83$ |
| CB700M2 | R = 1073<br>C = 0.047<br>A = $2.05 \times 10^{-3}$<br>$\alpha = 0.8$ | $s' \approx 1.05$;<br>$s'' \approx 1.83$ | $s' \approx 1.05$;<br>$s'' \approx 1.85$ | $s' \approx 1.07$;<br>$s'' \approx 1.81$ |
| CB700M | $s' \approx 1.05$;<br>$s'' \approx 1.83$ | $s' \approx 1.04$;<br>$s'' \approx 1.85$ | $s' \approx 1.03$;<br>$s'' \approx 1.88$ | $s' \approx 1.04$;<br>$s'' \approx 1.84$ |

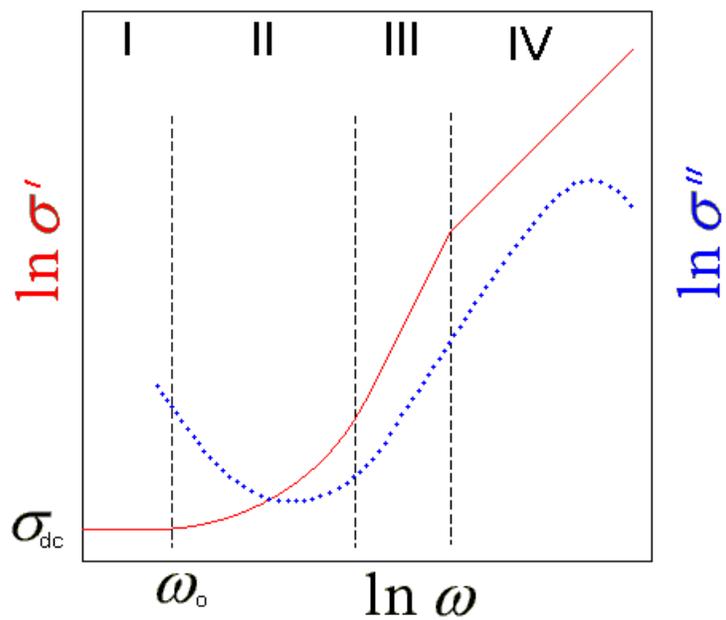

FIG. 1. General conductivity spectra for boron doped amorphous carbon films. The solid curve (——) is for real part of conductivity and the dotted curve (·······) for imaginary part of conductivity.

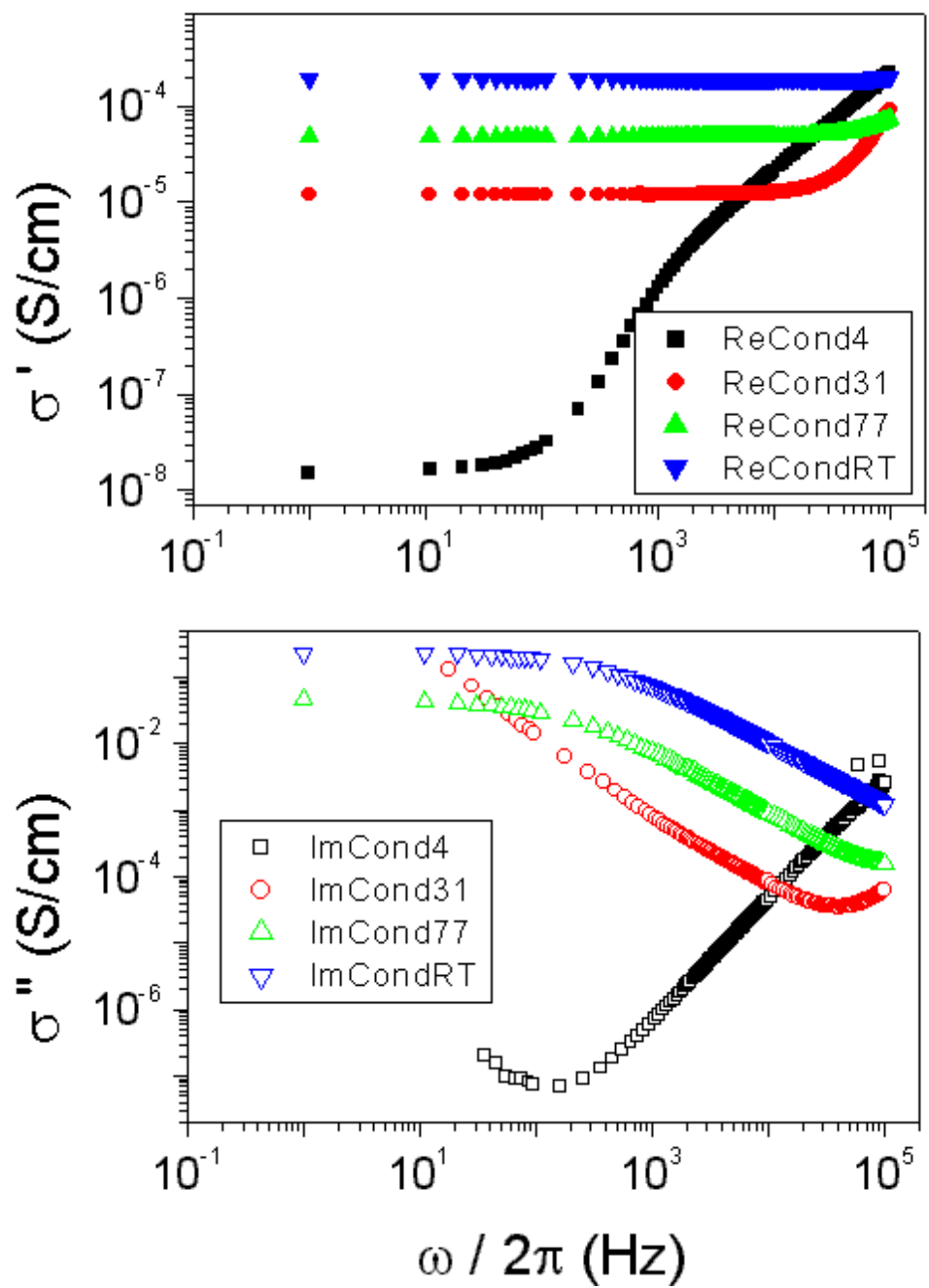

FIG. 2. Conductivity spectra for C700 films at 4.2K(■), 31K (●), 77K (▲) and 300K (▼). The top figure with solid symbol is for real part of conductivity and the bottom figure with open symbol for corresponding imaginary part of conductivity.

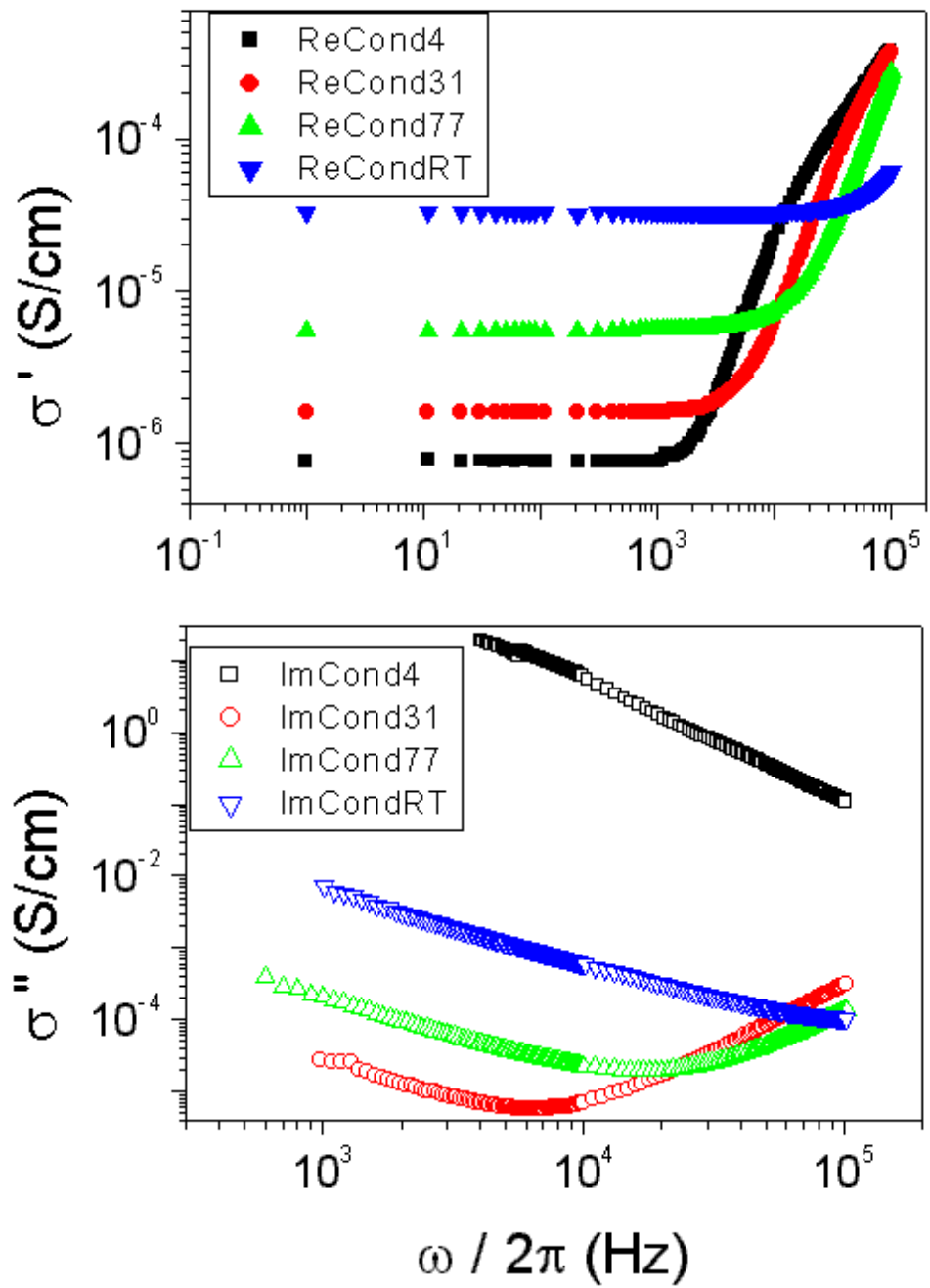

FIG. 3. Conductivity spectra for CB700M4 films at 4.2K(■), 31K (●), 77K (▲) and 300K (▼).The top figure with solid symbol is for real part of conductivity and the bottom figure with open symbol for imaginary part of conductivity.

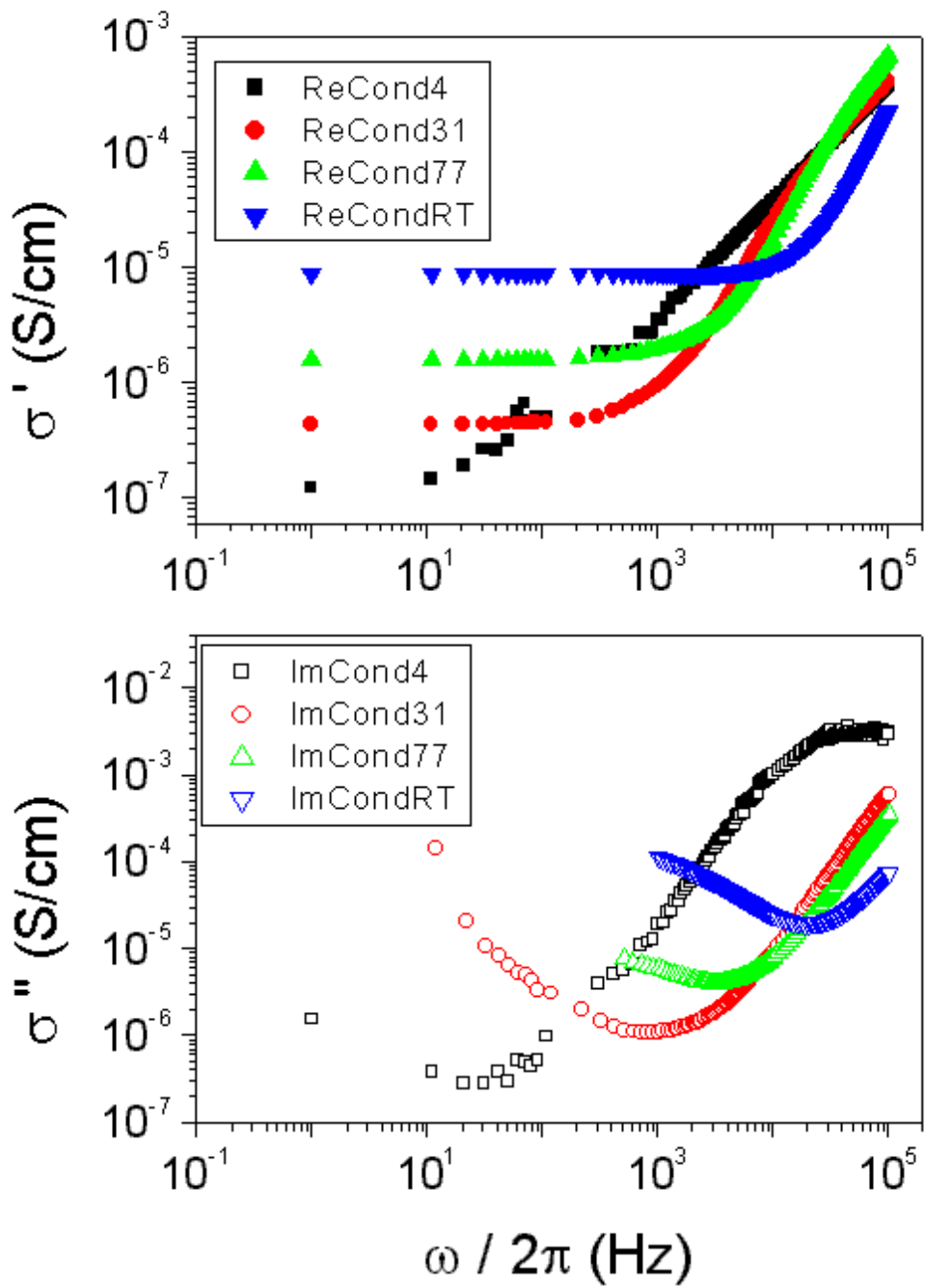

FIG. 4. Conductivity spectra for CB700M2 films at 4.2K(■), 31K (●), 77K (▲) and 300K (▼). The top figure with solid symbol is for real part of conductivity and the bottom figure with open symbol for imaginary part of conductivity.

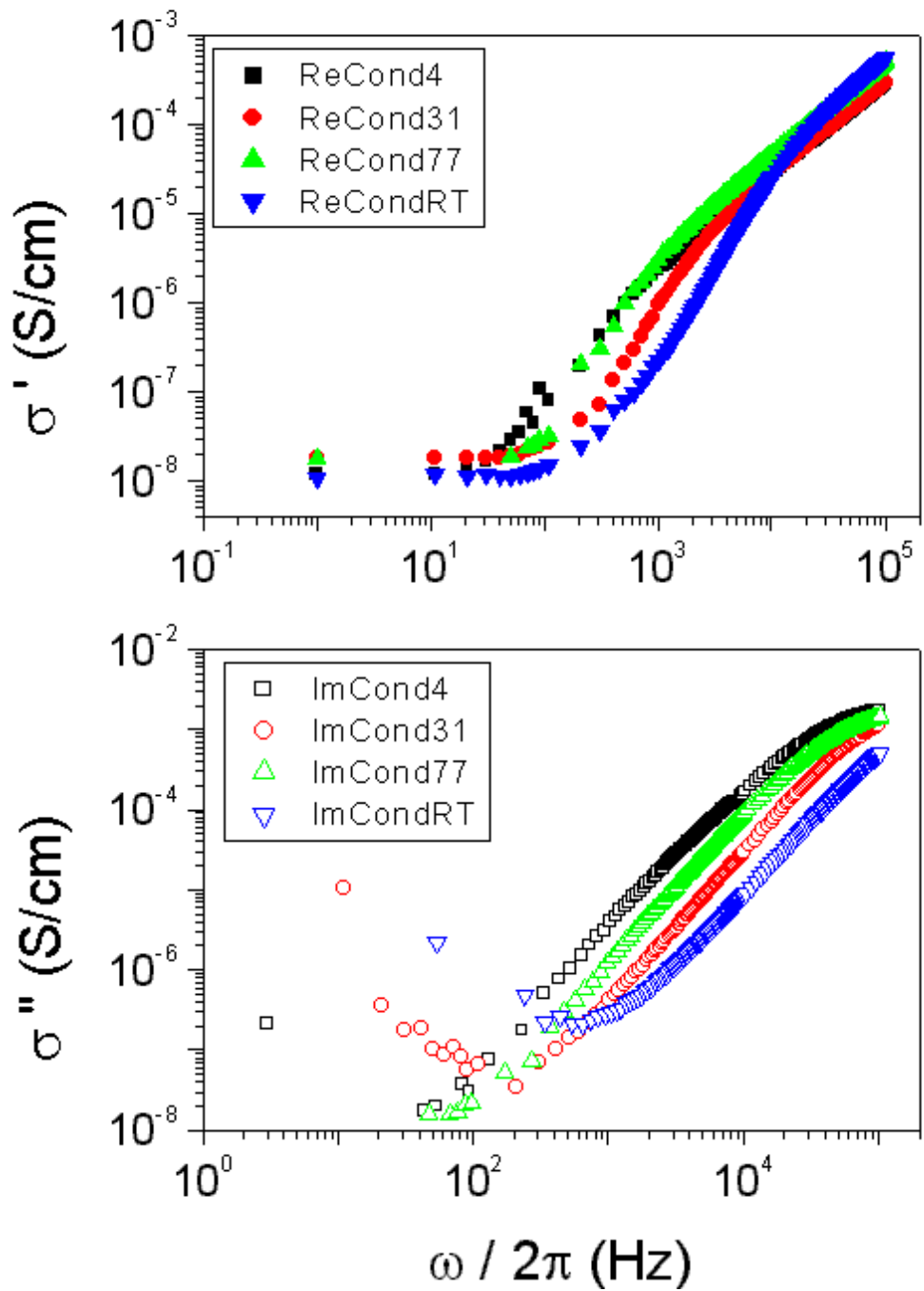

FIG. 5. Conductivity spectra for CB700M films at 4.2K(■), 31K (●), 77K (▲) and 300K (▼). The top figure with solid symbol is for real part of conductivity and the bottom figure with open symbol for imaginary part of conductivity.

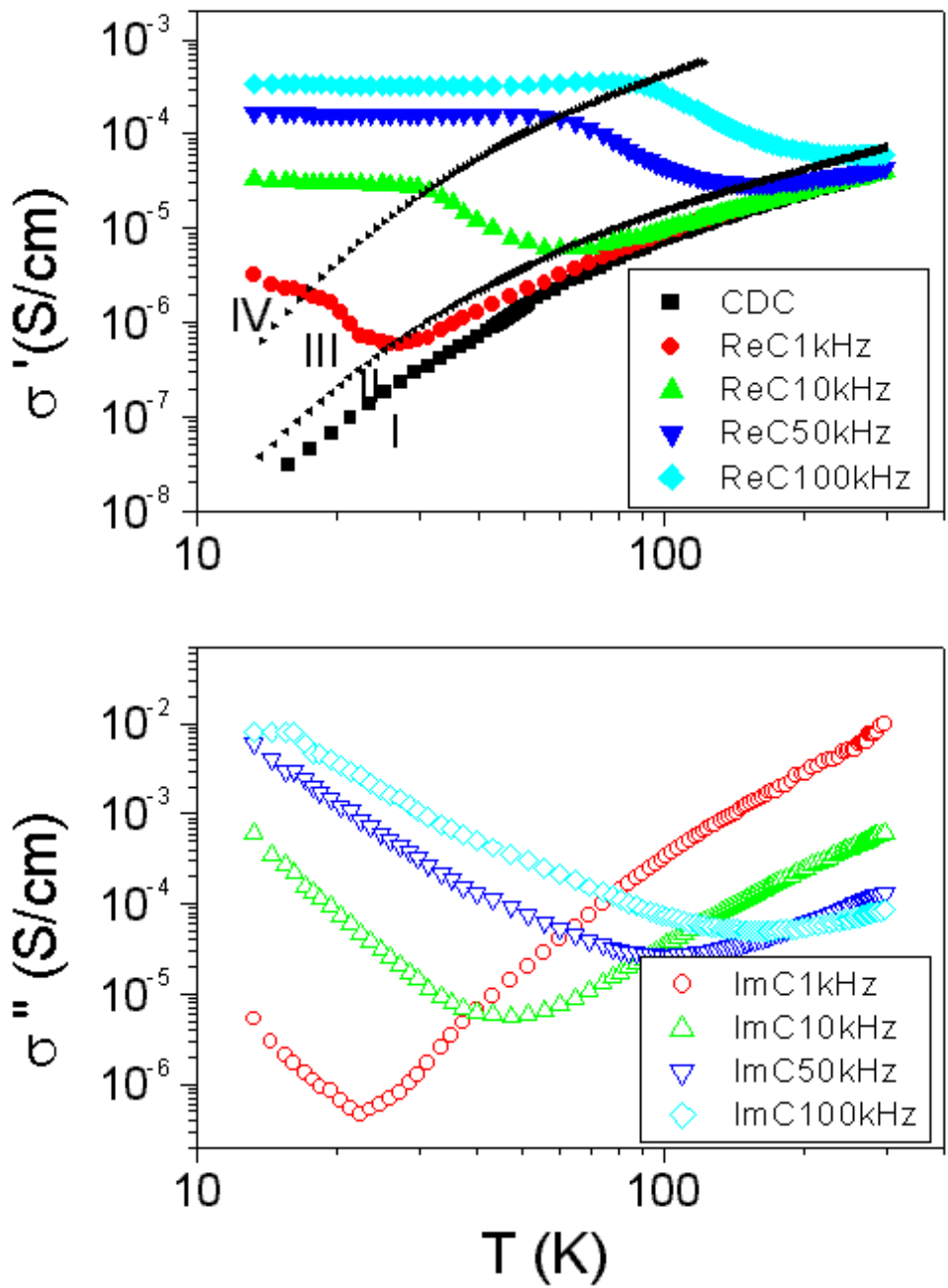

FIG. 6. The temperature dependence of ac conductivity for C700 at 1kHz (●), 10kHz (▲), 50kHz (▼) and 100kHz (◆) along with dc conductivity (■). The dotted lines and the dc conductivity curve divide the entire spectra in four regimes. The top figure with solid symbols is for real conductivity and the bottom one with open symbols is for imaginary conductivity.

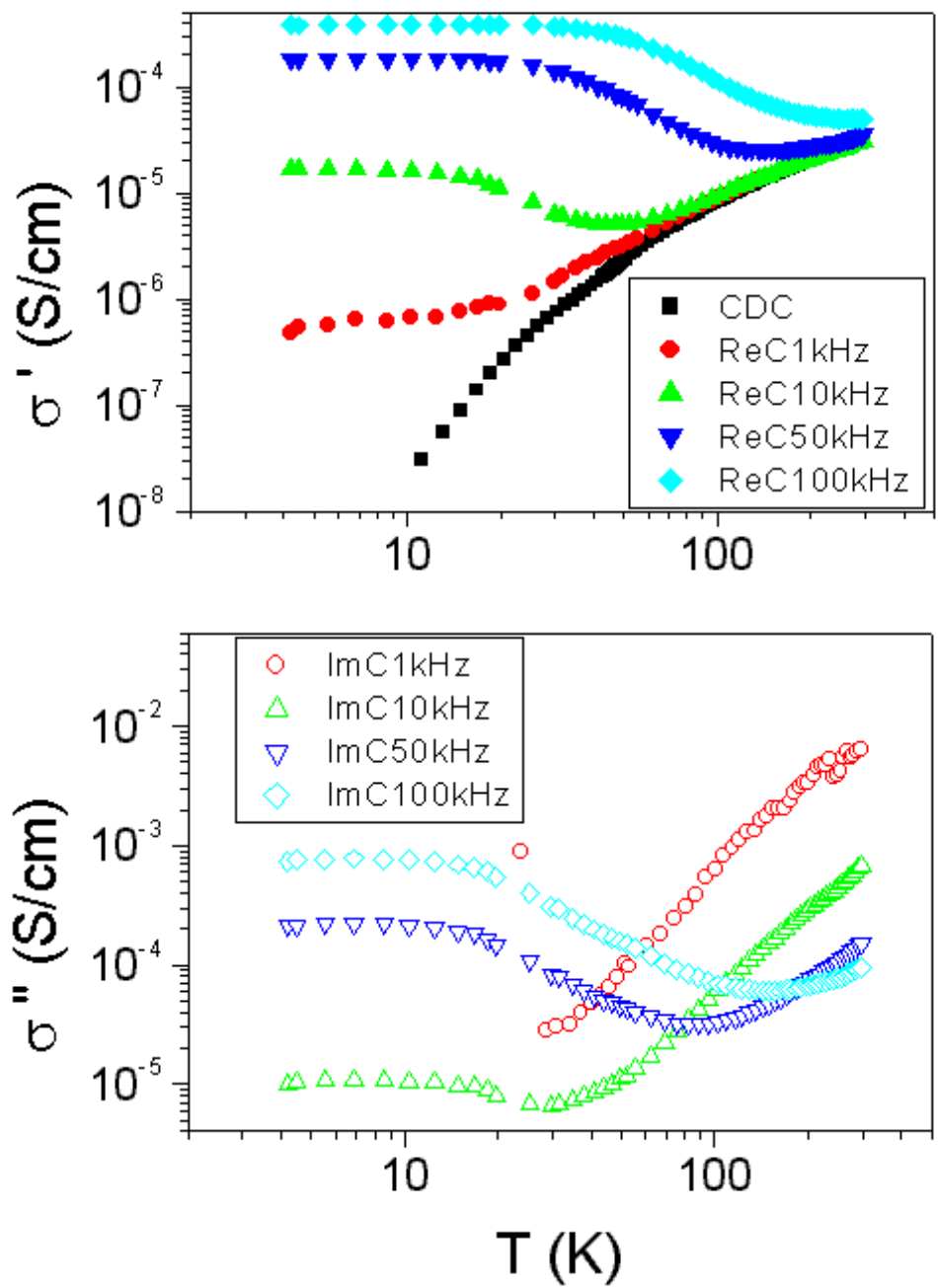

FIG. 7. The temperature dependence of ac conductivity for CB700M4 at 1kHz (●), 10kHz (▲), 50kHz (▼) and 100kHz (◆) along with dc conductivity (■). The top figure with solid symbols is for real conductivity and the bottom with open symbols is for imaginary conductivity.

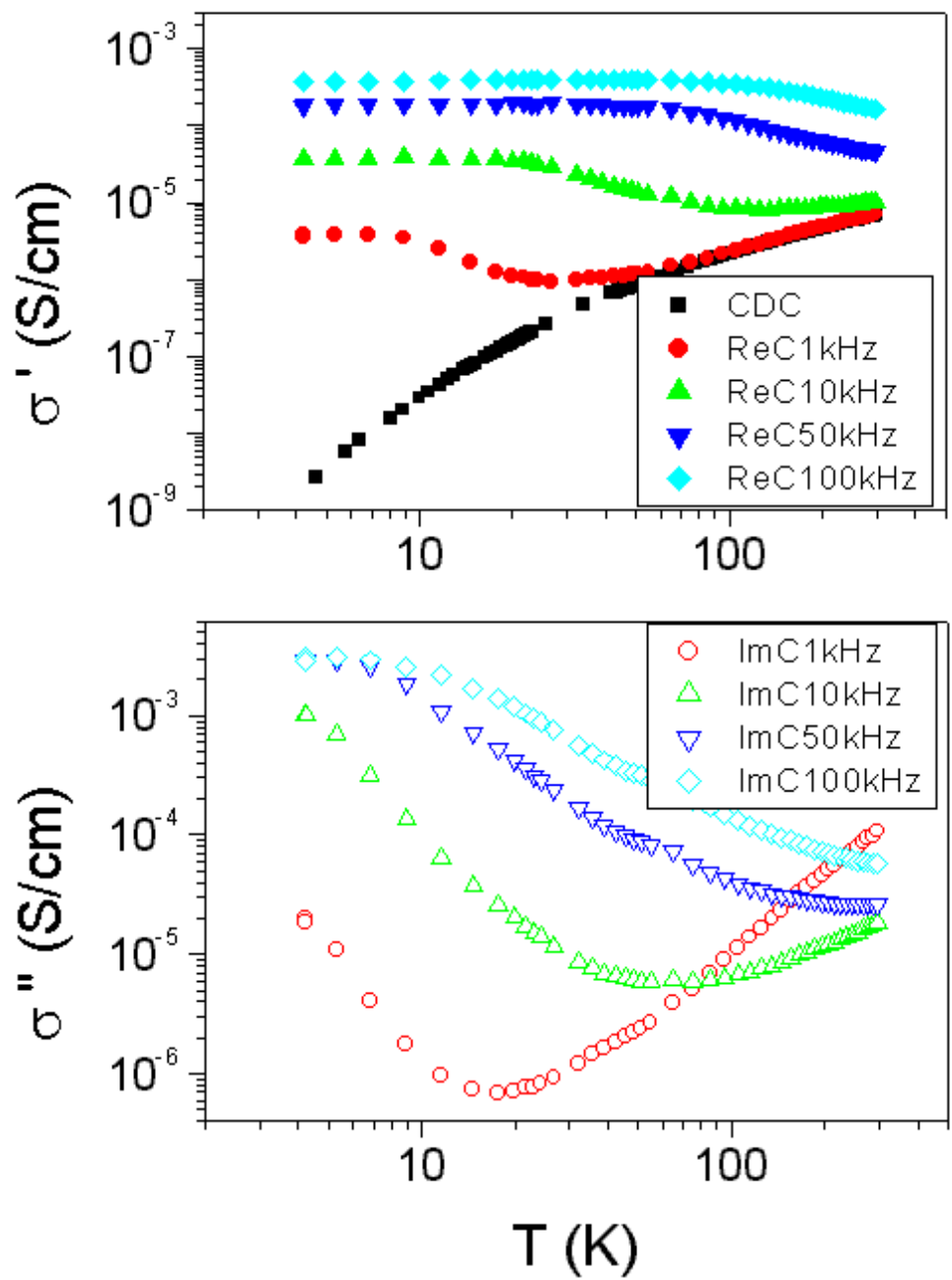

FIG. 8. The temperature dependence of ac conductivity for CB700M2 at 1kHz (●),
10kHz (▲), 50kHz (▼) and 100kHz (◆) along with dc conductivity (■). The top figure
with solid symbols is for real conductivity and the bottom one with open symbols is for
imaginary conductivity.

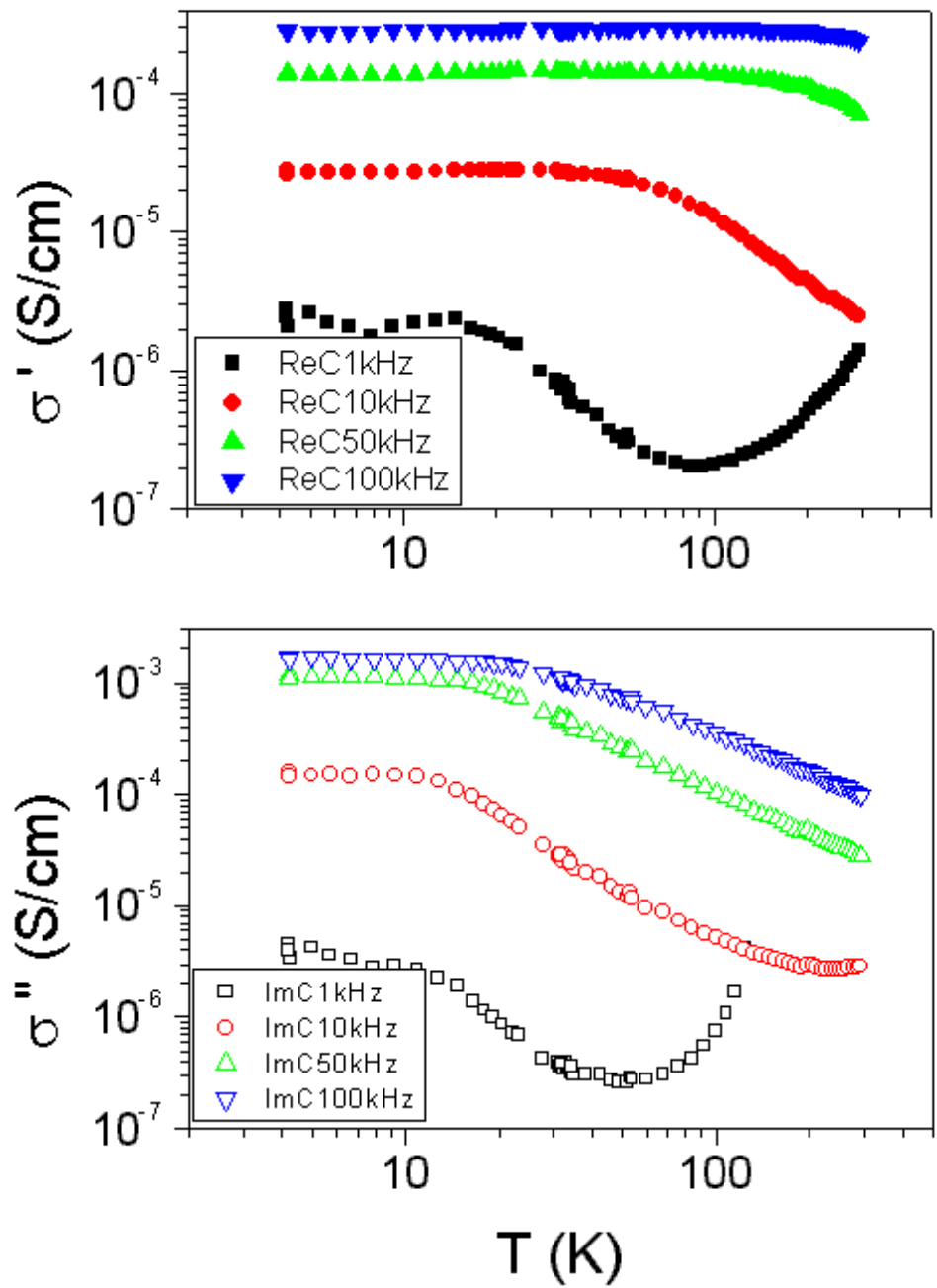

FIG. 9. Temperature dependence of ac conductivity for CB700M at 1kHz (■), 10kHz
(●), 50kHz (▲) and 100kHz (▼). The top figure with solid symbols is for real
conductivity and the bottom one with open symbols is for imaginary conductivity.